\documentclass[12pt]{article}

\usepackage[margin=1in]{geometry}
\usepackage{amsmath,amssymb,graphicx,color}
\usepackage[titletoc,title]{appendix}
\usepackage{bm}
\usepackage{mathrsfs}
\usepackage{authblk}

\title{From identification of random contributions to determination of the optimum forecast of a soccer match }

 \author{Andreas Heuer}
 \affil{University Muenster, Institute of Physical Chemistry, 48149 Muenster, Germany}

\begin{document}
\maketitle

\begin{abstract}
The forecasting of sports events is of broad interest from the applied but also from the theoretical perspective. In this work the question is addressed for the example of the German soccer Bundesliga how a theoretically optimum forecast of the goal difference of a match can be characterized. This involves a careful analysis of the random contributions in a match and its disentanglement from the informative contributions, resulting from the individual team strengths. An important aspect is the consideration of the time dependence of the team strength which turns out to mainly fluctuate around a team-specific value during the course of a season. Two types of time-dependent properties have to be distinguished, one being uncorrelated between different match-days, the other being correlated and thus accessible by an appropriate correlation analysis. For some performance indicators, which may be used to estimate the team strength, the quality of the respective forecast is compared to the theoretical optimum. Knowledge of the informative contribution allows one to conclude that the offensive team strength is more important than the defensive team strength for the final success.
\end{abstract}

\section{Introduction}

Forecasting deals with the question how to make the best out of the available data. Typically, one is faced with two conceptually different types of statistical errors when forecasting the outcome of a future experiment from the knowledge about past experiments. Firstly, the available data set from the past is finite which automatically generates some uncertainty. Secondly, thinking, e.g., of coin tossing, even if the knowledge about the coin is perfect, the outcome of a future experiment (e.g., the number of heads after 10 tosses) contains a random component. Both uncertainties add up in the evaluation of the forecasting quality.

When talking about the minimum uncertainty, corresponding to the optimum forecast, one has take into account that via a more profound analysis of previous measurements it might be possible to obtain additional information. For the case of a coin one might not only observe the results from the available experiments but also analyse via some physical methods, e.g, its shape to identify possible deviations from a fair coin. Thus, for a truly optimum forecast one should take into account that in principle the system is characterized in full detail.

The key goal of this work is to discuss the optimum forecast of a soccer match. Like for coin tossing, a perfect forecast of the result of a specific match is not possible. Indeed, it was shown that the final result can be approximately
described via uncorrelated Poisson processes \cite{Maher82,janke1,janke2}. Subsequently, a more
detailed analysis has revealed that the actual number of draws is estimated as too small by this approach \cite{Riedl15}.  This effect can be explained in terms of the loss aversion as expressed by the prospect theory \cite{Kahneman79}, i.e. via underlying psychological effects. In general, these random effects reduce correlations of performance indicators with external input, such as signing a new contract
 \cite{Gom19}. Fluctuations due to random effects are also reported for  match-to-match \cite{Liu16} or in-match variations \cite{Rampinini09}. It also explains why the underdog has a small but significant chance to win a match \cite{ben3}.

 In analogy to the example of coin tossing, knowledge about previous results to estimate the outcome of future matches must be taken with care because random effects in these results may bias the estimation process. In particular, taking the goals from previous matches for forecasting purposes,  one may expect that the random contributions are very high due to the small number of goals in a soccer match. Thus, older approaches, based on previous match results \cite{Lee97,Dixon97,Dixon98,Rue00} have a limited predictability \cite{Link16}.  For other performance indicators, displaying a larger number of occurrences (chances for goals, number of passes, ...) it is likely that the random effects are less disturbing \cite{Link16,Tenga10, Memmert17,heuer_buch}.  In recent years, the advent of tracking data has improved the quantitative description of different aspects
of soccer matches; see, e.g., for a review \cite{Sarmento18}. Based on tracking data one can get even more involved observables. One example is the so-called packing rate \cite{internet3}. It is related to the number of passes. In contrast to the mere number of passes, the individual passes are evaluated together with appropriately weighting factors, expressing the success of the passing behavior \cite{Memmert17}.

Of key interest is the optimum forecasting of the goal difference. However, it may be also very instructive to forecast other observables such as the chances for goals or the number of passes. In this way new insight about the nature of soccer matches can be gained. This analysis complements previous work where it was analysed how well an observable can predict the team strength \cite{heuerarxiv,Heuer14}. As a side effect the statistical framework allows one to identify the variation between different teams after removing random effects. We apply this information to the question whether the offensive or the defensive strength of teams are more important for the final success.

The outline of this work is as follows. The data, used for this work, are introduced in Section II. The statistical background is presented in Section III. Key ideas are discussed for the simple case of coin tossing and, subsequently, the results are applied to the case of the forecasting of soccer matches.   Section IV contains the analysis of the random and informative contributions, as applied to different observables. In Section V the implications for the optimum forecasting are discussed. We conclude and summarize in Section VI.

\section{Data basis}

From www.kicker.de we took the goals and the chances for goals for
all matches from the season 2005/06 to the season 2016/17 for the
German soccer Bundesliga. Specific criteria are used to define a chance for goals. They
are consistently applied by journalists from that sports journal. In previous work \cite{Heuer14} it has been shown that
from a statistical perspective the scoring of a goal corresponds to a random process such that with probability of approx. 25 \%
a chance for a goal is transformed into a goal. This factor of 0.25 is within very small variations ($\pm 0.01$) nearly identical for all teams.  Furthermore, we compare observables which are related to passes.
First, for the seasons 2009/10 and 2010/11 we took the number of passes from
www.kicker.de. Second, we obtained the packing rate from IMPECT \cite{internet5} for the two seasons 2015/16 and 2016/17. Going beyond a simple counting of passes, it weights each pass with the number of opponent players who are taken out of the match. For  the specific definition of the packing rate one analysed all passes
and dribbles  during a match; see also \cite{internet4}. Let before the pass have $P$ players from the opponent standing closer to
the opponents goal than the ball. Whenever after a pass (or dribble) the ball is still controlled by the ball
possessing team, one compares the values of $P$ before and after the pass (dribble). If
$P$(before) is larger than  $P$(after), the pass (or dribble) has bypassed
$\Delta P = P$(before) - $P$(after) $> 0$ players. To determine the packing rate one adds up $\Delta P$ for all corresponding events in a match. Safe passes to the back of the
field (which formally would yield $\Delta P < 0$) thus do not contribute.

\section{Statistical background}

\subsection{Tossing a coin}

For the discussion of the statistical framework we consider a variable $X$ which is characterized by a distribution with variance $\sigma^2$.  We start by drawing $N$ different values $S_i$ ($i \in \{ 1,...,N \}$) of that variable and denote them as a {\it strength}. Additionally we require that the average of these $N$ strength values is exactly zero. As an example one might think of $N$ different coins with different probabilities $p_i$ to toss a head (+1) or a tail (-1). Its strength is then defined as $S_i = 2p_i - 1$ such that a fair coin would have a strength of zero. A single measurement $x_i$ comprises a finite number of throws and subsequent averaging of the outcome, resulting in a number between -1 and +1. For later purposes we introduce the notation $\langle x_i \rangle$ which describes the measurement in case of an infinite number of throws, i.e. where all random effects have been removed. It is, of course, identical to the strength $S_i$.

Additionally, we consider $M$ measurements of each coin, which we write as $x_{i}(t) = S_i + \delta_{i}(t)$. Here $\delta_{i}(t) $ is a random number and $t \in \{ 1,...,M \}$. We assume that all random numbers are drawn independently from a common distribution, characterized by the variance $\delta^2$. This variance can be interpreted as measurement noise.

Now we randomly take $n \le M$ measurements $x_{i}(t)$ for fixed $i$ and take
their average value. Later it will become important that this subset is randomly chosen and does not correspond to, e.g., just the first $n$ measurements. This analysis is repeated for each $i$. The variance of that distribution is denoted  $V(X,n)$.
Under the assumptions, formulated above, one obtains from elementary statistics
\begin{equation}
\label{varxn}
V(X,n) = \sigma^2 + \frac{1}{n} \delta^2.
\end{equation}
Thus, the variance has two contributions. The first term reflects the variance of the $S_i$ without any measurement noise and the second term expresses the measurement noise, weighted by the factor $1/n$ (less noise for more measurements). We denote $\sigma^2$ the {\it informative}, and $\delta^2$ the {\it random} contribution.

Naturally, for an infinite number of measurements ($n=\infty$) one could easily obtain the informative contribution, i.e. $\sigma^2$. However, even for finite number of measurements $M$, Eq.\ref{varxn} can be used to obtain $\sigma^2$ and $\delta^2$ separately when having data for at least two different values of $n$. Technically, a better method is to determine $V(X,n)$ for all accessible values of $n$ ($1 \le n \le M$) and to perform a linear regression analysis of $V(X,n)$ vs. $1/n$.

In a more general setting the strength may be time dependent, i.e., described by $S_i(t)$ such that $x_i(t) = S_i(t) + \delta_i(t)$. The time-dependence of the strength can be conceptually decomposed into two contributions
\begin{equation}
S_i(t) = S_i + \kappa_{i,0}(t) + \kappa_{i,1}(t)
\end{equation}
with random numbers $\kappa_{i,0}(t)$ and $\kappa_{i,1}(t)$ characterized by variances $\kappa_0^2$ and $\kappa_1^2$, respectively. The first term $\kappa_{i,0}(t)$ shall reflect random contributions which are uncorrelated in time whereas $\kappa_{i,1}(t)$ contains the random contributions which are correlated in time. For the modelling we choose the simple process
\begin{equation}
\kappa_{i,1}(t+1) = a \kappa_{i,1}(t) + \eta_i(t)
\end{equation}
with a random number $\eta(t)$, uncorrelated in time and with variance $\eta^2$. Straightforward calculation yields for the time correlation function (introducing $\Delta S_i(t) = S_i(t) - S_i $)
\begin{equation}
\label{timecorr}
f(\Delta t) = \langle \Delta S_i(t) \Delta S_i (t + \Delta t) \rangle = \kappa_1^2 \exp(-\Delta t / t_0) ,
\end{equation}
i.e., an exponential decay with the decay time $t_0 = -1/\log(a)$ and variance $\kappa_1^2 = \eta^2/(1 - a^2)$. The average is over all times $t$ and all coins $i$. In what follows we always assume that $1 \ll  t_0 \ll M$. Please not that $f(\Delta t)$ is insensitive to the random contributions $\kappa_{i,0}(t)$ which are uncorrelated in time.

Taking into account the time-dependence of the strength values, the random contribution increases and  Eq.\ref{varxn} is generalized to
\begin{equation}
\label{varxn_new}
V(X,n) = \sigma^2 + \frac{1}{n} [ \delta^2 + \kappa_0^2 + \kappa_1^2] \equiv \sigma^2 + \frac{1}{n} V_{tot}.
\end{equation}

Now $\sigma^2$ reflects the variance of the distribution of the average strength values $S_i$. There is one subtle aspect with Eq.\ref{varxn_new}. The $1/n$ scaling only holds, if the random numbers of the different measurements are uncorrelated. On first sight this might be at odds with the fact that $\kappa_{i,1}(t)$ is correlated in time. However, since the $n$ data points are randomly taken out of the $M$ measurements (see above) and since  $t_0 \ll M$, these correlations are not relevant and Eq.\ref{varxn_new} is, indeed, fully justified.

\subsection{Soccer matches}

Now we specify the analysis for the case when the variable $X$ is identified with the goal difference $\Delta G$. First, we concentrate on a single season. Then $x_{i}(t)$ can be identified with the goal difference from the perspective of team $i$ on match day $t$.  The value of $N$ denotes the number of teams in a league ($N=18$ for the German soccer Bundesliga), the number $M$ can be as large as the number of match days ($M = 2N-2 = 34$).  However, as compared to the above analysis there is the complication that the match result not only depends on the quality of team $i$ but also on that of the opponent (denoted as team $j$) in that match. In what follows, the goal difference in that match of team $i$ vs. team $j$ is denoted $g_{ij}$ (which, in this example, can be identified with $x_{i}(t)$).

It is well known, that there exists a home advantage such that the goal difference $g_{ij}$, averaged over all matches, is positive (this average is denoted $g_{home}$). Interestingly, it turns out via some appropriate statistical analysis (see, e.g., \cite{Heuer09}) that to an excellent approximation the expected home advantage is constant for all matches within a season. Thus, we determine $g_{home}$ for a given season and substitute $g_{ij}$ by $g_{ij} - g_{home}$ before performing the statistical analysis. In this way we get rid of side effects due to the home advantage. The same procedure is used for the other performance indicators, studied in this work.

In analogy to above,  $\langle g_{ij} \rangle$ denotes the expectation value of the goal difference of a match $i$ vs. $j$ for which all random contributions are taken out. As already discussed above, this can be conceptually defined as an average over an infinite large number of realizations of the same match under identical conditions. Next we define the team strength $S_i$ of a team $i$ as the average of $\langle g_{ij} \rangle$ over all matches of team $i$ in a given season. This additional average over the different teams is denoted as $[.]_{j; j \ne i}$.
Thus one has
\begin{equation}
S_i =[ \langle g_{ij} \rangle ]_{j; j\ne i}.
\end{equation}
This is more complicated than in the case of the coin and just reflects the fact that a specific match depends on two rather than a single team strength. 
In practice, the individual values of $S_i$ can only be estimated due to the random contributions fo the actual goal differences $g_{ij}$. However, this is the same situation in the case of throwing a coin where just from knowledge of the outcome of a finite number of throws it is not possible to predict the exact strength.

This definition has been already employed in previous work \cite{Heuer09,Heuer10}.  By construction one has $[ S_i]_i = 0$. The variance of the distribution of team strengths, i.e. $[ S_i^2]_i$  is denoted $\sigma^2_S$. By construction $S_i$ reflects the quality of team $i$, averaged over the whole season. An average team has $S_i \approx 0$. We first consider the case where the team strength is constant during the season. In analogy to above we can directly generalize the results to the case of time-dependent team strengths.

Knowledge of the team strengths $S_i$ and $S_j$ allow one to calculate the expected goal difference $\langle g_{ij}\rangle $ in a match of $i$ vs. $j$ via \cite{Heuer10}
\begin{equation}
\label{eqgab}
\langle g_{ij}\rangle = \frac{N-1}{N} (S_i - S_j).
\end{equation}

Eq. \ref{eqgab} can be regarded as a first term in a Taylor-expansion, fulfilling the required symmetry criterion $\langle g_{ij}\rangle = - \langle g_{ji}\rangle$. The key result in Ref.\cite{Heuer10} was to show that already the next higher-order cubic term has no relevance when applied to real data. The prefactor $(N-1)/N$ has a simple origin. When averaging over the whole season when team $i$ has exactly played twice against all other teams $j$, one obtains for the expected goal difference of team $i$ the relation  $[\langle g_{ij}\rangle ]_{j;j \ne i} = [(N-1)/N](S_i - [ S_j ]_{j \ne i} )$. Since the sum over all team strengths must be zero,  one has the generally valid relation $ (N-1) [ S_j ]_{j \ne i} + S_i = 0$. This yields $[\langle g_{ij} \rangle]_{j \ne i} = S_i$, consistent with the definition $S_i$. To reach this self-consistency, the factor $(N-1)/N$ is essential

For later purposes we note that the team strength of two different teams $i$ and $j$ are anticorrelated. After a short calculation one obtains
\begin{equation}
\label{anticorr}
[ S_i S_j]_{i,j; i \ne j} = -\sigma_S^2/(N-1).
\end{equation}

Next we write in analogy to the case of throwing a coin
\begin{equation}
\label{grandom}
g_{ij} = \langle g_{ij}\rangle + \delta_{ij}
\end{equation}

 where $\delta_{ij}$ is the random contribution for a match characterized by a variance $\delta^2_{ij}$. The variance may indeed slightly depend on the specific teams.  If, e.g., both teams have a weak offensive but a strong defensive, one would expect a smaller number of goals which, within the Poisson approximation (see below for more details) would give rise to a smaller variance. We also introduce the average variance $\delta^2 = [\delta^2_{ij}]_{i,j; i \ne j}$. This is a key quantity, since $\delta^2$ characterizes the typical contribution to the goal difference of a match for which no forecast is possible.

Now we consider the distribution of the goal differences if averaging over $n$ ($1 \le n \le N-1$) match days. Its variance is denoted $V(X=\Delta G,n)$. In analogy to the case of coins we select the $n$ match days randomly from the whole season under the condition that each match $i$ vs. $j$  only appears once. Note that this condition restricts the allowed choices for the $n$ match days because in a round-robin tournament each match appears twice (e.g. when comparing the matches at match day 1 and $N$). Finally, one averages over the selection of the $n$ match days. In practice, we use 500 different choices. Finally, we average $V(X=\Delta G,n)$  over the different seasons.

For the evaluation of the statistical properties of these averaged goal differences, we write in generalization to Eq.\ref{varxn}
for the variance of the resulting distribution, using Eq.\ref{eqgab} and Eq.\ref{grandom},
\begin{equation}
V(X=\Delta G, n) = \left ( \frac{N-1}{N} \right )^2 \left [ \left ( \frac{1}{n} \sum_{j}^{}{}^\prime (S_i - S_j) \right )^2 \right ]_i + \frac{1}{n} \delta^2.
\end{equation}
The prime indicates that the sum is over $n$ opponents of team $i$. Naturally, one has $j \ne i$. By construction of the averaging procedure, all teams $j$ are different with respect to each other.

After evaluation of the squared term and using Eq.\ref{anticorr}, a straightforward calculation yields
\begin{eqnarray}
\label{varn2}
V(X=\Delta G, n) & = & \left ( \frac{N-1}{N} \right )^2 \sigma_S^2 \left ( 1 + \frac{2}{N-1} + \frac{1}{n^2}\left ( n - \frac{n(n-1)}{N-1} \right ) \right ) +\frac{1}{n} \delta^2 \nonumber \\ & = &  \sigma^2  +\frac{1}{n} (\sigma^2 +  \delta^2)
\end{eqnarray}
with
\begin{equation}
\label{sigma2}
\sigma^2 = \frac{N-1}{N} \sigma^2_S .
\end{equation}

Thus, one obtains a strict $1/n$-dependence, which allows one to determine $\sigma^2$ and $\delta^2$ individually. Note again that $\delta^2$ captures the random contributions to the final result whereas $\sigma^2$ expresses the informative contributions to the resulting goal difference of a match.

The term $\sigma^2/n$, which did not appear in Eq.\ref{varxn}, expresses the fact that the cumulated goal difference of team $i$ becomes less dependent on the team strengths of the opponents after averaging over more and more matches. For its derivation we had to introduce the formal concept of the team strength.

For a better understanding of Eq.\ref{varn2} we may easily reproduce the two limits $n=1$ (distribution of goal differences after one match day) and $n=N-1$ (distribution of goal differences after half of the season). In the first case one has $V(X=\Delta G,n=1) = ((N-1)/N)^2 [(S_i -S_j)^2]_{i,j;i\ne j} + \delta^2 = ((N-1)/N)^2 \sigma_S^2 (1 + 1 + 2/(N-1)) + \delta^2 = 2 \sigma^2 + \delta^2$ . In the second case $n=N-1$ one can use the fact that $\sum_j S_j = -S_i$ so that $V(X=\Delta G,n=N-1) = ((N-1)/N)^2 [(S_i +S_i/(N-1) )^2]_{i} = \sigma_S^2 + \delta^2/(N-1) = (N/(N-1))\sigma^2 + \delta^2/(N-1)$. Both results agree with the general expression Eq.\ref{varn2}.

In analogy to above this analysis can be generalized to the case of time-dependent team strengths $S_i(t)$. In this case Eq. \ref{varn2} generalizes to
\begin{equation}
\label{varn2_new}
V(X=\Delta G, n)  = \sigma^2  +\frac{1}{n} (\sigma^2 + V_{tot})
\end{equation}
with
\begin{equation}
V_{tot} =  \delta^2 + 2 \kappa_0^2 + 2\kappa_1^2.
\end{equation}
 in analogy to Eq.\ref{varxn_new}. The factor of 2 reflects the dependence of the goal difference on the team strengths of both teams.

  \begin{figure}[h!]
  \includegraphics[width=0.8\textwidth]{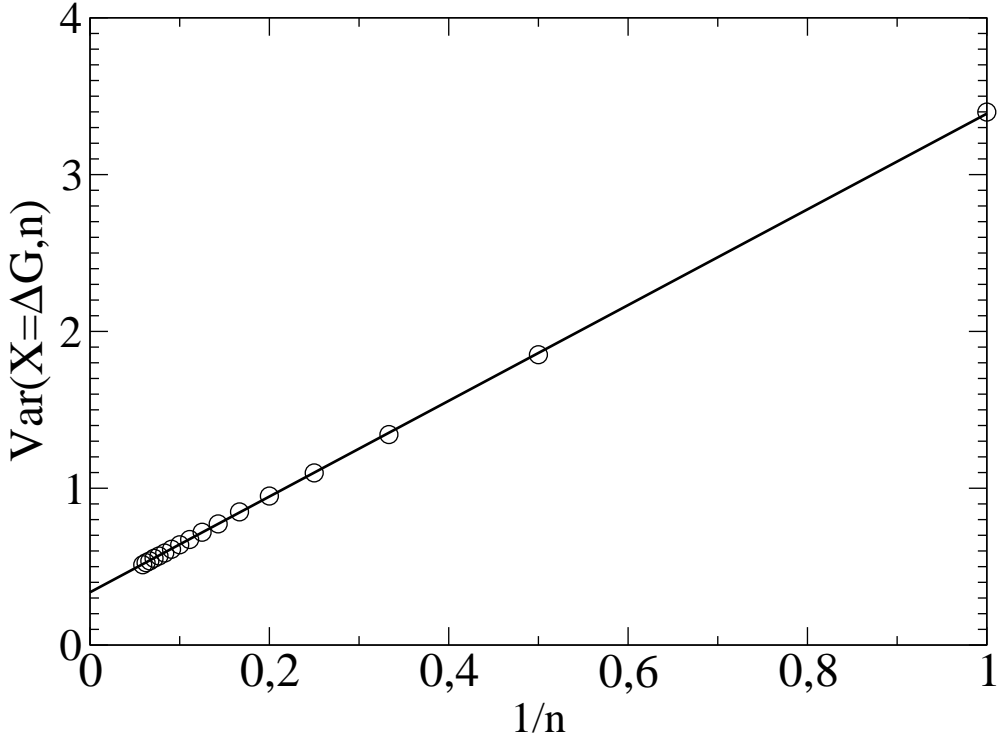}
  \caption{The $1/n$-dependence of the variance $V(X,n)$ for the example of the goal difference $\Delta G$. The straight line is the result of a linear regression analysis, yielding the values $V_{tot} = 2.71$ and $\sigma^2 = 0.34$.}
      \end{figure}

To exemplify the statistical analysis, we show in Fig.1 the $1/n$-dependence of $V(X=\Delta G,n)$. Here we have averaged $V(X=\Delta G, n)$ over all 12 seasons in order to obtain a good statistics, expressing the average behavior during this time interval. One observes a perfect linear relation, yielding the values $\delta^2 + \kappa_0^2 + \kappa_1^2 = 2.71$ and $\sigma^2 = 0.34$.

Naturally, this analysis can be also used for other observables beyond the goal difference in order to disentangle the informative and random effects of different variables, relevant in soccer matches. If we take other differences (e.g. passes of team $i$ minus passes of team $j$) the practical as well as the conceptual analysis is fully analogous.
The situation is a little more complex, if one just takes, e.g., the number of goals $g_i$ scored by team $i$ in a match against team $j$. This number depends on the offensive strength of team $i$ and the defensive strength of the opponent $j$. This involves two different distributions, characterizing the distribution of offensive strengths and that of defensive strengths. However, the numerically determined goals of the individual teams, averaged over $n$ match days (in full analogy to above), display again a $1/n$ behavior. Extrapolation of the limit of large $n$ allows one to again to read off the variance of the distribution of offensive team strengths. It is denoted $\sigma^2(X_+)$ with $X=G$. Similarly, taking as input information the goals, conceded by a team, one obtains the distribution of defensive team strengths $\sigma^2(X_-)$. These two values will be also analysed further below.

Finally, we note that for the error analysis we have constructed a set of synthetic results with exactly the statistical features as found in this work. By repeated simulation of these models and using exactly the same procedure as for the actual data one can directly read off the statistical errors.

\subsection{Optimal forecasting / individual random contributions}

Let us assume that $M$ matches are played and one wants to predict a specific match of $i$ vs $j$ at match day $M+1$. Optimum forecasting implies that one {\it exactly} knows the team strengths $S_i(t=M+1)$ and $S_j(t=M+1)$. How well a specific observables is suited for estimating the team strength has been discussed in \cite{heuerarxiv}, but this is beyond the scope of the present work. If $S_i(t=M+1)$ and $S_j(t=M+1)$ are known, the resulting uncertainty is just given by $V_{opt} = \delta^2$. Here we assume that the variance $\delta^2$ is identical for all teams but, of course, this could be easily generalized. Three important aspects have to be taken into account.

(1) It may be extremely difficult to estimate the team strength at a given match if the team strength depends on time. For the correlated fluctuations (characterized by the variance $\kappa_1^2$) one might take information from the previous matches to estimate the present deviation from the average behavior of a team. However, even if $S_i(M)$ would be exactly known, there is still some uncertainty with respect to $S_i(M+1)$. If the correlation time $\tau_0$ becomes large these uncertainties become small. Therefore we require for the optimum forecast that impact of the time-correlated fluctuation is known.  In case of uncorrelated fluctuations (characterized by $\kappa_0^2$) no previous information is available at all. Thus, one might as well characterize  the optimum forecasting by the variance $V_{opt} = \delta^2 + 2 \kappa_0^2$. However, in principle it is conceivable that specific information just before the match (e.g., the line-up of the teams and thus the information about injured players) may help to estimate to the contribution of the uncorrelated fluctuation to the team strength. In any event, it is a matter of taste whether the term $2 \kappa_0^2$ is taken into account when expressing the optimum forecasting. In what follows, we will always choose $V_{opt} = \delta^2 + 2 \kappa_0^2$.

(2) The variance $\kappa_1^2$ can be fortunately directly extracted from the time correlation function Eq. \ref{timecorr}. Then after determination of $V_{tot}$ via Eq.\ref{varn2_new} one can directly determine $V_{opt} = V_{tot} - 2 \kappa_1^2$. Please not that a priori it is not possible to individually determine $\delta^2$ and $\kappa_0^2$.

(3) The situation becomes more simple if one has an underlying model for the estimation of the uncertainty of a measurement. For the example of the coin one knows that for given strength one can estimate $\delta^2$ based on knowledge of the binomial distribution. As shown for the case of soccer the distribution of goals follows with very high accuracy from a Poissonian distribution \cite{heuer_buch} which allow one to estimate $\delta^2$. Therefore, we can also explore the knowledge of $\delta^2 + 2\kappa_0^2 + 2\kappa_1^2$ together with the estimation of $\kappa_1^2$ and $\delta^2$  to estimate the match-to-match fluctuations as characterized by $\kappa_0^2$.

\section{Results}

\subsection{Time correlation}

The goal is to characterize the time dependence of $S_i(t)$.  A priori it is not clear how big the typical deviations from the average value $S_i$ are.
A preliminary analysis of this question can be found in
\cite{Heuer09}. There it has been shown that after dividing the
whole season in four parts of (nearly) equal length, the
correlations among each pair of quarters is identical within the
statistical error. The average team strength only varies when a
new season starts \cite{Heuer09}.

For a more quantitative analysis we estimate the time correlation function $f(t)$ as defined in Eq.\ref{timecorr}. To calculate this function we might, in principle, multiply the goal difference of team $i$ at match day $t$ with the goal difference at match day $t + \Delta t$.
Since the goal difference is basically given $S_i - S_j$ plus some random value, in the product only the product $S_i(t) S_i(t + \Delta t)$ will remain since the team strengths of the two opponents as well as the random contributions are averaging out. Of course, this would not hold for $\Delta t = 17$ because then the two opponents would be identical. Therefore $\Delta t = 17$ has to be omitted from the analysis. Here we neglect the small effects, resulting from the finite number of teams in a season.

Even when taking into account 16 different seasons the statistical fluctuations of $f(t)$ would be extremely high due to the large random contributions when analysing the goals. It has been proven extremely helpful to study the chances for goals. As discussed in Ref.\cite{Heuer14}  this
observable fully reflects the team strength but with much better
statistical properties (less random effects) than the number of
goals.

  \begin{figure}[h!]
  \includegraphics[width=0.8\textwidth]{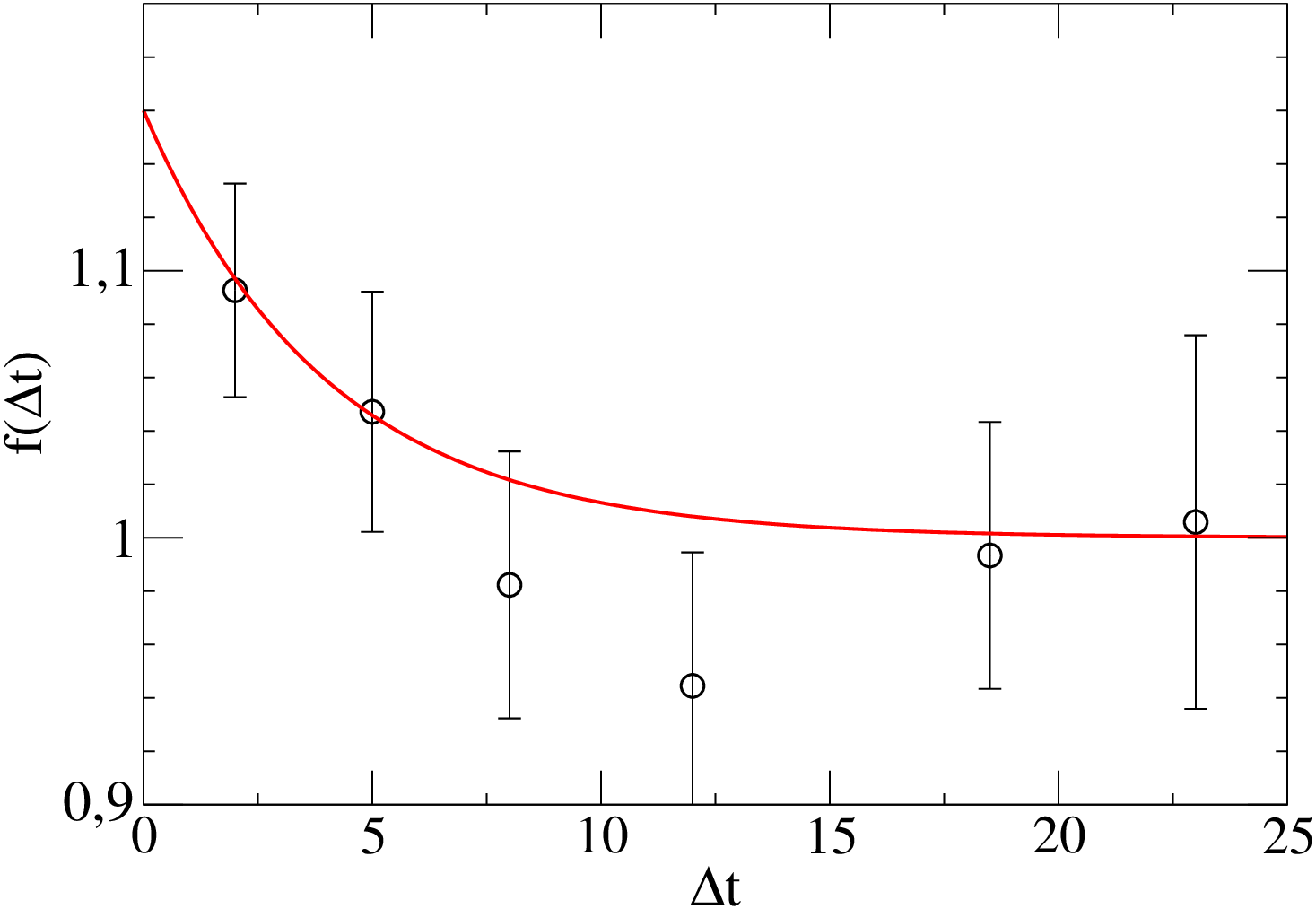}
  \caption{The autocorrelation function $f (\Delta t)$
  for the difference of the chances for goals as a measure of the
  team strength. It is averaged over the 16 seasons  2001/02
until 2016/17 in order to improve the statistics. The long-time limit is normalized to
  unity.
   Included is an exponential fit of the form
    $ 1 + c_1 \exp(-\Delta t / c2)$ with   $c_1 = 0.16$ and $c_2 = 4.0$. Fits
    of similar quality are possible for $3 \le c_2 \le 6$.}
      \end{figure}

The result is shown in Fig.2.  To first
approximation, the team strength is constant during the season
since the autocorrelation function hardly changes with $\Delta t$.
A closer view shows that the team strength varies on the time
scale of approx. 4 match days. Note that this time scale is much
smaller than the 17 match days, corresponding to half of the
season. Furthermore, describing the short-time behavior of $f(\Delta t)$ by an additional exponential function (see figure caption of Fig.2) we find that the variance of the fluctuations is significant but relatively small (16\% with respect to $\sigma_S^2$). In the above terminology this implies $\kappa_1^2 = 0.16 \sigma_S^2 $. Together with the relation  $ \sigma_S^2 = (18/17) \sigma^2$ this yields
\begin{equation}
\label{kap1}
\kappa_1^2 = 0.17 \sigma^2.
\end{equation}
Inserting values yields  $\kappa_1^2 = 0.17 \cdot 0.34 = 0.058 \approx 0.24^2$.
 For the time being we neglect the possible contributions of $\kappa_0^2$ (which, anyhow, will turn out to be very small). Then,  within 68\% probability the team
strength will fluctuate in an interval $[-0.24, 0.24]$ around the
respective average team strength. Based on the properties of the correlation function and using a time constant $c_2 = 4$
we have simulated a stochastic process with Gaussian noise and a
Gaussian distribution of team strengths which agrees with these properties. The results are shown in
Fig.3. This figure may help to get an idea about the origin of the short-time decay
of the correlation function in Fig.2. For this specific realization of team strengths the best team is somewhat separated from the rest. This corresponds to the actual reality since a closer analysis reveals that the distribution of team strengths is described by a somewhat skewed Gaussian-type distribution (data not shown, see also \cite{heuer_buch}).

  \begin{figure}[h!]
 \includegraphics[width=0.9\textwidth]{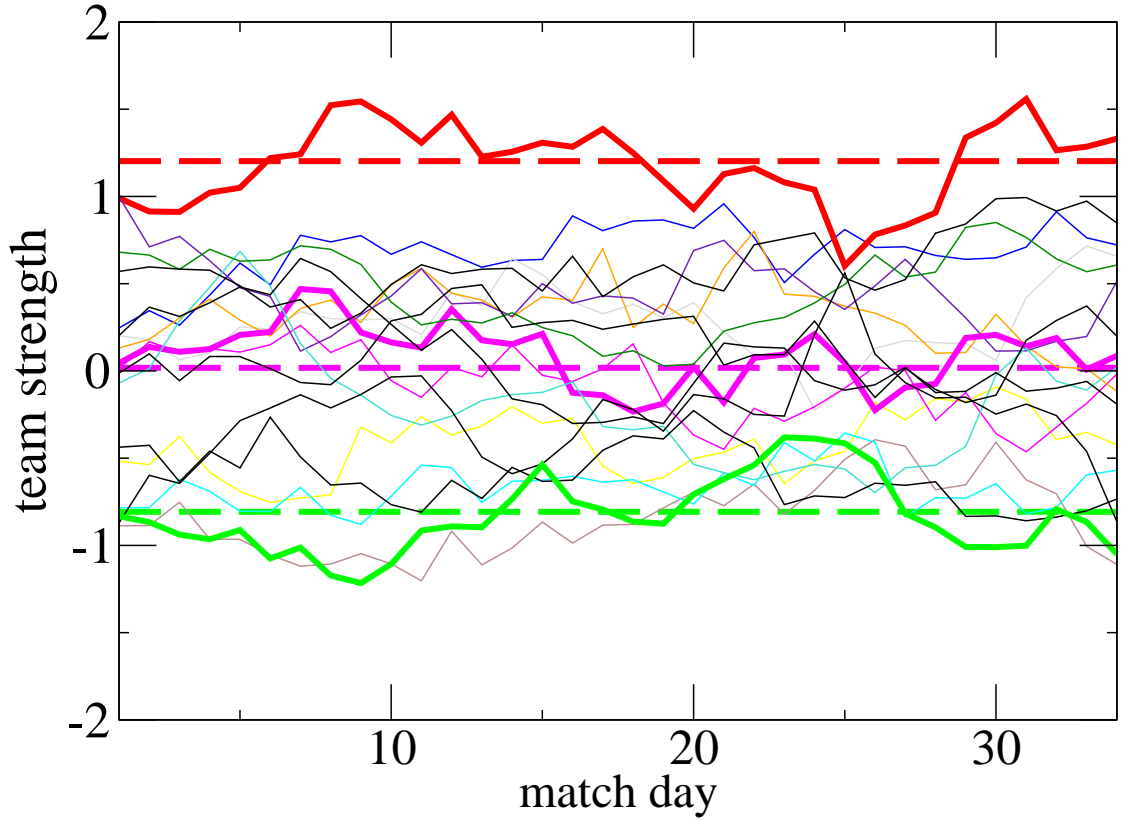}
  \caption{Stochastic simulation of the time dependence of the team strength of 18 teams, incorporating
  the distribution of team strengths as well as the fluctuations in a quantitative manner. A few examples (best team, average team, worst team) are
  highlighted. The
  broken lines reflect the average team strengths of these teams.}
      \end{figure}

\subsection{Properties of random contributions}

The distribution of the sum of the random contributions are characterized by the variance $V_{tot}$. We aim for determining also the individual contributions, i.e. the values $\delta^2$, $\kappa_0^2$, and $\kappa_1^2$. For the specific case of goals it expresses the contribution to the final result of a match which cannot be forecasted by the season-averaged values of the team fitness of a team. Naturally, the variance $\delta^2$ can also be determined for other observables. Here we particularly consider the case of goals, chances for goals, the packing rate,  and the number of passes.  In analogy to above we always consider the differences such as, e.g., the goal difference in a match.  We start with the discussion of the goal difference and then discuss in analogy the other observables.

For convenience we list all statistical parameters which are used for the subsequent discussion in Tab.2.

\begin{table}[h!]
\caption{The different statistical parameters, characterizing the modelling of the variance. $V_{tot}$ and $\sigma^2$ are derived from the $1/n$-fit as outlined above. $2 \kappa_1^2$  is always chosen as $ 2 \cdot 0.16 \sigma^2  \sigma^2$. $N_{match}$ denotes the average number of the observable $X$ per match. For the case of packing $f \ne 1$ expresses the assumed average packing rate per event. $\delta^2$ expresses the random contribution for known team strengths (see text for the estimation based on $N_{match}$ and $f$).  The last column contains the estimated value of $ \kappa_0^2 = (V_{tot} - \delta^2)/2 - \kappa_1^2$.}
\label{tab2}
      \begin{tabular}{l|l|lllll|l}
        \hline
         X  & $V_{tot}$  & $\sigma^2$ & $2 \kappa_1^2$  &  $N_{match}$ & $f$ & $\delta^2$  & $ \kappa_0^2$  \\ \hline
       Goals & 2.71 $\pm 0.06$ & 0.34 & 0.11 & 2.86 & 1 &2.61 & 0 \\
         Chances & 15.1 $\pm 0.2$ & 3.03 & 1.03 & 11.36& 1 & 11.36 & 1.3 \\
       Packing & 3430 $\pm 150$   & 4230  & 1440 & 566 & 2 & 1130 & 430\\
        Passes & 11260 $\pm 330$ & 4910 & 1670  & 588 & 1 & 588 & 4500\\ \hline
      \end{tabular}
\end{table}

{\it Goal difference:} From the $1/n$-fit we obtain $V_{tot} = 2.71$, from Eq. \ref{kap1}  the value of $2\kappa_1^2 = 0.11$. This results in $V_{opt} = \delta^2 + \kappa_0^2 = 2.60$. This is the minimum uncertainty when forecasting an (average) soccer match in the Bundesliga (when the time-correlated variation of the team strength is fully taken into account).

One can go one step further and try to rationalize the value of $V_{opt}$. As summarized in the Introduction, the scoring of goals is to a good approximation a Poissonian process. To check this we have along the lines of Ref.\cite{Riedl15,heuer_buch,Heuer14} predicted the team strengths for the teams, based on the chances for goals of the other 33 match days in that season. The result is shown in Fig.4a. Indeed, for the home as well as for the away teams one observes an excellent agreement. However, forecasting the goal difference of a match, using the assumption of independent Poissonian processes, one obtains deviations for small goal differences; see Fig.4b. The actual data have too many draws. On a qualitative level this is compatible with previous results shown in \cite{Riedl15,heuer_buch} and explained in that work based on psychological effects.  If the two Poissonian processes were independent, the variance $\delta^2$ would be given by the average total number of goals in a match $N_{match} = 2.86$ (when averaging over all teams and seasons). Taking into account the correlation effect, increasing the number of draws, the resulting variance is reduced. Indeed, based on the data in \cite{Riedl15} this reduction can be estimated to be approx. 0.25. We note in passing that this reduction effect was significantly larger in earlier times when a victory was awarded by 2 rather than 3 points \cite{Riedl15}. Thus, the modified Poissonian estimation gives rise to a variance to $2.86 - 0.25 = 2.61$ which is basically identical to the value of $V_{opt} = 2.60$, estimated above.

  \begin{figure}[h!]
  \includegraphics[width=0.45\textwidth]{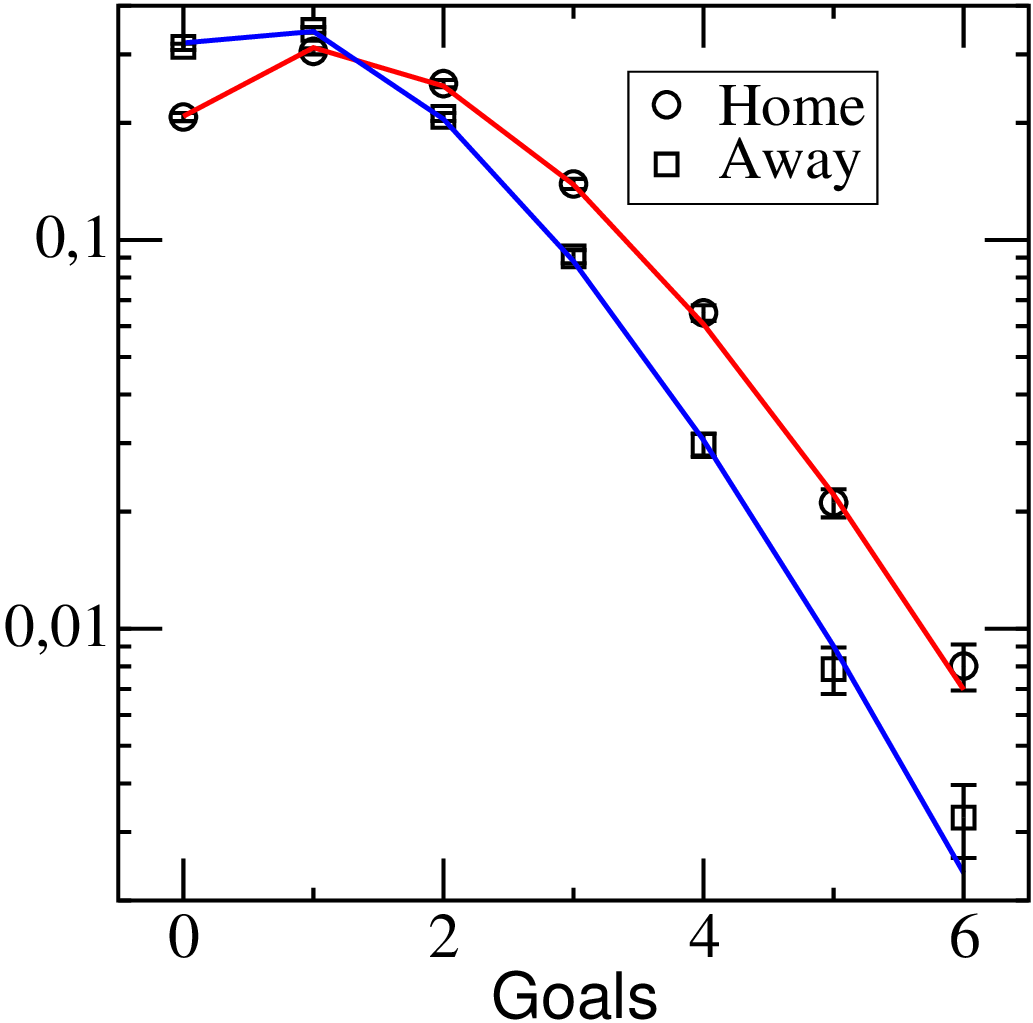}
    \includegraphics[width=0.45\textwidth]{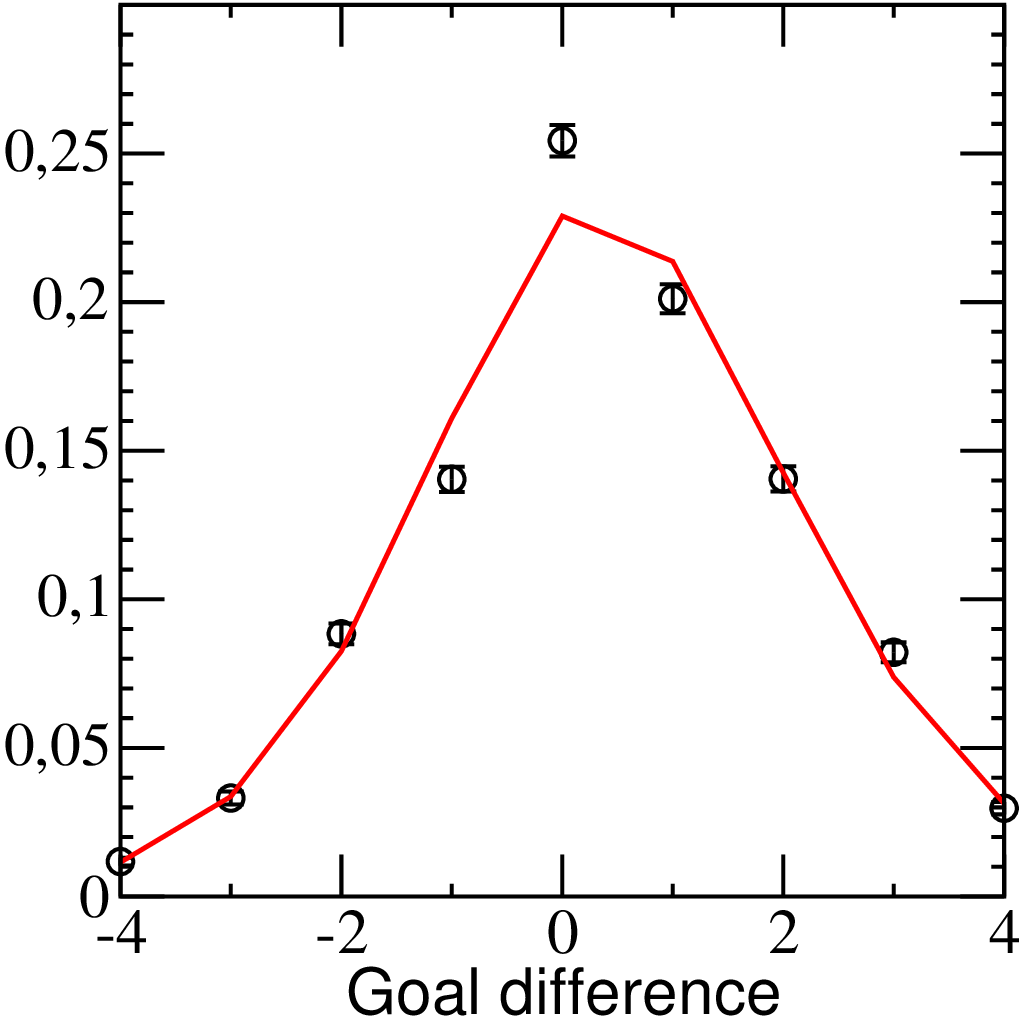}
  \caption{Left: The  number of goals, scored by the home and the away team, respectively, based on the seasons 1995/96 until 2016/17. The symbols denote the empirical data, the solid lines the Poisson predictions, based on an estimation of the team strength (see text).  Right: The distribution of goal differences from the perspective of the home team. For the Poisson predictions independent scoring of goals is assumed.}
  \end{figure}

We would like to stress that the determination of $V_{opt}$ is just data-based and without any underlying model assumptions. In contrast, the comparison with the Poissonian approach is somewhat more subtle. The value of 2.86 for independent Poissonian processes, based on the number of goals, is well-defined and just follows from the total number of goals. The precise value for the reduction due to correlations is based on the comparison of independent Poissionian distributions and the actual data. For the first distribution the estimation of the team strengths was essential, naturally involving some estimation error. However, since the probability of a draw hardly depends on the difference of the team strengths in a match \cite{heuer_buch}, the estimation of the reduction of the variance by 0.25, mentioned above, should not be influenced too much by possible estimation errors.

As a consequence of these arguments the modified Poissonian contribution to the random effects in a soccer match exclusively reflect the variance $\delta^2$ since minor uncertainties in the values of the team strengths would not change that estimate. Based on the values, mentioned so far, the remaining variance $\kappa_0^2$ is very close to zero, i.e. the time-dependent fluctuations of the team strength mainly result from correlated effects, reflecting, e.g., the fact that injuries or psychologically different times typically take longer than a single week.

{\it Chances for goals:} In analogy to the goals one can ask the question whether it is possible to forecast for a specific match the difference of the chances for goals. Of course, this forecast if of no relevance for, e.g., the betting of soccer matches. However, as discussed below, one can learn a lot about the mechanisms of a soccer match. The analysis is fully analogous to the case of goals. There is just one difference: since the presence of an identical number of chances for goals of both teams has no psychological implication, we neglect any possible deviations from a purely Poissonian behavior. Furthermore, we assume that Eq.\ref{kap1} is always valid (actually, as discussed above, in Fig.2 we had determined this ratio based on the chances for goals). We find $\delta^2 \approx 11.4$ and $\kappa_0^2 \approx 1.3$. Note that again $\kappa_0^2 $ is much smaller than $\delta^2$ but significantly larger than 0.  There are two possible interpretations. First, this might indicate that the ability to generate chances for goals (or to void chances for goals of the opponent) display a match specific characteristics. However, since we did not observe them for the team strength, defined for the resulting goals, and due to the strong relation between goals and chances for goals \cite{Heuer14} this hypothesis is  unlikely. Rather it might indicate that the definition of the chances for goals involves, despite strict criteria, some intuition of the journalists. If this amounts to have a single chance for goals more or less in a match, this additional contribution of approx. 1  can be easily rationalized.

{\it Packing rate:} Next we would like to discuss the packing rate with respect to its
random aspects. Again we take the difference in analogy to the goal difference.
Given the strong relation between team strength and packing rate, observed in \cite{heuerarxiv}, we proceed analogously. From the $1/n$-analysis we obtain $V_{tot} \approx 0.8 \sigma^2$. Note that in contrast to the goals and chances for goals the informative contributions are larger than the random contributions. Partly this is just a consequence of the fact that there are more events in a match which contribute to the packing rate.

When estimating the Poissonian contribution we assume that each  event, contributing to the packing rate, is contributing with a value of 2. Indeed this value is larger than unity, see, e.g., \cite{Memmert17}, although a value of 2 is just a rough estimation. The actual choice does not modify any of the conclusions, drawn below.
Thus, a packing rate of 566 (see Tab.\ref{tab2}) corresponds to approx.
283 events. Assuming a Poissonian distribution, the variance of
the number of events is also 283 and thus the variance of the
packing rate is $2^2 \cdot 283 \approx 1130$. This may be considered as an upper limit because in practice there may be minimum time intervals between different successful events with high packing rates. Using the data, shown in Tab.1, one ends up with $\kappa_0^2 \approx 430$. This value is nearly one order of magnitude smaller than $\sigma^2$. Thus, as for the goals and the chances for goals the match-to-match fluctuations are also relatively small with respect  for the packing rate. This is compatible with the observation that the packing rate reflects the team strength very well. Thus, the absence of major match-to-match fluctuations for the team strength suggests a similar behavior for the packing rate.

{\it Passes:} We have also performed this analysis for the number of passes of a
team. The main difference to the packing rate is the dramatically increased variance $\kappa_0^2$ which is close to $\sigma^2$.   These results suggest that with respect
to the number of passes one has significant match-to-match fluctuations,
reflecting, e.g., the different tactics in that match.  For example, in a specific match the team manager may choose the tactics to play very defensive with little ball possession and thus a small number of passes. Naturally, this shows up as a very large match-to-match fluctuations. Since passes are much less correlated with the team strength as the packing rate \cite{heuerarxiv} this result is not at odds with the results for $\kappa_0^2$, discussed so far.

\subsection{Relevance of offensive vs. defensive actions}

There is another interesting aspect which can be derived from knowledge of the informative contributions. Let us consider a virtual league
where during a match of $i$ vs. $j$ each team has a fixed number of
shots at a goal wall. The final result expresses the number of
successful shots of both teams. If we perform the analogous
statistical analysis for this example, one finds a major
difference to real soccer matches. There is nothing like a
defensive strength because the number of hits of team $i$ is not
influenced by any properties of team $j$ and vice versa.

More generally, we may ask to which degree the actions, performed
by the ball possessing team, are influenced by the defensive
strength of the opponent. The answer to this question can be
quantified by comparing the values for the variances
$\sigma^2(X_+)$ and $\sigma^2(X_-)$ of the respective performance indicators $X_+$ and $X_-$ , obtained by application of
Eq.\ref{varxn}. In the extreme case that the outcome of a match is
only influenced by the offensive strength (as for the case
described above) one would expect $\sigma^2(X_-) = 0$ because,
formally speaking, the defensive strength is identical (namely
totally irrelevant) for all teams.

In Tab.\ref{tab4} we show the values $\sigma^2(X_\pm)$ for goals and
chances for goals. Let's assume that team $i$ is
playing vs. team $j$. In all cases, $X_+$ reflects a variable which
can only vary when team $i$ possesses the ball. If $\sigma^2(X_-) >
0 $ it is possible for team $j$ to influence the success of team $i$
via its defensive strength. In general, the ratio
$\sigma^2(X_+)/\sigma^2(X_-)$ is a direct measure of the relevance
of the offensive strength relative to the defensive strength with
respect to the variable $X$. These values are also listed in
Tab.\ref{tab4}.

\begin{table}[h!]
\caption{The noise-free variances of different performance indicators. The final column reflects the relevance of the
offensive strength as compared to the defensive strength.}
\label{tab4}
      \begin{tabular}{llll}
        \hline
         X  & $\sigma^2(X_+)$  & $\sigma^2(X_-)$ & $\sigma^2(X_+)/\sigma^2(X_-)$   \\ \hline
       Goals & 0.155 $\pm 0.012$ & 0.07 $\pm 0.009$&2.2 $\pm 0.3$\\
        Chances for goals & 1.56 $\pm 0.076$   & 0.86 $\pm 0.058$ & 1.8 $\pm 0.15$ \\
      \hline
      \end{tabular}
\end{table}

We find that for goals and chances for goals this ratio is approx.
2. This means that the offensive strength is more important than
the defensive strength. For the specific case of goals, this
result (albeit with an approximative extrapolation scheme for the
determination of $\sigma^2(X_\pm)$) was already mentioned in
Ref.\cite{Heuer09}. This means that both offensive and defensive
strengths are important although the offensive contribution is
even more relevant. This indicates a limited (but still very
relevant) impact of defenders to stand against excellent strikers.

\section{Implications for the quality of forecasting}

From the results, discussed so far, interesting insight about the forecasting of soccer matches can be extracted. Here we concentrate on the forecasting of the goal difference. Of course, all arguments can be easily generalized to different observables, relevant, e.g., in the context of betting. In particular, the present results allow one to quantify some aspects where so far only a qualitative understanding was possible.

The first aspect, to be discussed, deals with a simple question: to which degree can a soccer match be forecasted at all? For this purpose one can compare the variance of the goal difference, one obtains without random effects (e.g., by having fictive matches of very long duration) with the actual variance. The fraction $f$ of the final result, which is not due to random effects can thus be written as
\begin{equation}
f = \frac{2\sigma^2 + 2 \kappa_1^2 + 2 \kappa_0^2}{2\sigma^2 + V_{tot}}.
\end{equation}
With the data from Tab.1 we obtain f=23\%. If one would considered the fluctuations of the team strength as a random effect, corresponding to omitting the $\kappa_i^2$-terms in the numerator, one would end up with f=20\%. In any event, the part of the soccer match which can be predicted, is quite small.
 Of course, for a match between two teams of identical
team strength the outcome is 100\% non-predictable whereas in a
match of a very good and a very bad team a larger fraction than
20\% is predictable.

Let us assume that 17 matches have been played and one wants to forecast the goal difference during the second half of the season. In this case the possible fluctuations of the team strength have to be interpreted as a random effect. Thus, the uncertainty per match is given by $V_{tot} = 2.71$.  This means that after 17 matches the uncertainty (in terms of the standard deviation) is $(2.71\cdot 17)^{0.5} = 6.8$. Thus, a forecasting with a smaller error is not possible. Of course, in practice the deviations would be higher because the estimation of the team strength will always have statistical uncertainties (see also below).

One may perform the same analysis for the points, using the standard 3-point rule (3 points for a win, 1 point for a draw, no point for a loss). In analogy to Fig.1, the data $V(X={\rm Points},n)$ are perfectly linear in $1/n$.  One obtains from a regression procedure $V(X={\rm Points},n)= 1.55/n + 0.12$. Using Eq.\ref{varn2}, the value $V_{tot}$  can be estimated as 1.55 - 0.12 = 1.43. In analogy, the uncertainty in estimated the points, gained by a team during the second half, is given by $(1.43\cdot 17)^{0.5} = 4.9$. If (hypothetically) the team strength could be already determined at the beginning of the season, the uncertainty in forecasting the points, gained during the whole season would be $4.9\cdot 2^0.5 = 6.9$. 

There have been attempts to predict the final points, e.g., in \cite{Cintia15}. For the season 2013/14 of the German soccer league they obtained a standard deviation of 9.1 based on using a pass-related performance indicator. Thus, this estimation is (in terms of variances) a factor of 1.7 (=(9.1/6.9)$^2$) worse than the theoretical optimum, averaged over the last 12 years. Actually, restricting our statistical analysis to this single season we basically get (by chance) the same value of $V_{tot}$ as if averaging over 12 years, as discussed so far. Most likely the reason for this deviation is the fact that the estimation of the team strength based on pass-related information is quite poor \cite{heuerarxiv}. The authors of that work argued that teams with strong  deviations were either very efficient/inefficient in scoring goals from goal attempts or the opponents displayed the inverse efficiency (in total captured by the so-called Pezzali score \cite{Cintia15}). Taking this effect into account the estimation could be significantly improved (probably even better than the above theoretical limit). However, this argument has to be taken with care. By taking the Pezzali score one uses explicit knowledge about the efficiency of a team to score goals or to avoid conceding goals in that season. However, the resulting bias in the goal difference is trivially correlated with the number of points. As a consequence, one can no longer speak of a forecasting procedure.

Finally, it may be instructive to check for explicit data how well the forecasting would really work for the goal difference of the second half of the season, based on the information from the first half. Let us denote the variance of this uncertainty as $\lambda^2$. Naturally, we expect that $(\lambda^2)^{0.5} > 6.8$.  For this purpose we rely on the information, taken from \cite{heuerarxiv}. We need to know how well the team strength can be estimated. We write
\begin{equation}\label{estS}
S_{i,est} =  \sqrt{1 - \frac{\epsilon^2}{\sigma_S^2}} S_i + \epsilon_i
\end{equation}
where the variance of the $\epsilon_i$-distribution is denoted $\epsilon^2$. The factor in front of $S_i$ guarantees that the variance of the estimated team strength is identical to the actual variance. Then the Pearson correlation of the distribution of $S_{i,est}$ and $S_i$ is given by $(1- \epsilon^2/\sigma_S^2)^{0.5}$. The values for this Pearson correlation (denoted $A_X$) are listed in Tab.3 in Ref.\cite{heuerarxiv} and are reproduced here in Tab.3.  In a first step this allows one to calculate the variance $\epsilon^2$. We also add the case where no information is added, i.e. the values of the strengths are randomly taken out of a distribution with variance $\sigma_S^2$. In this case one formally starts with $A_X = 0$.
Then we can express the variance $\lambda^2$ of the final uncertainty of the estimated goal difference as
\begin{equation}
\lambda^2 =  17^2 (17/18)\epsilon^2 + 17 V_{tot}.
\end{equation}
The factor (17/18) takes into account that we have first to transfer the variance of the team strength to the variance of the resulting goal differences according to Eq.\ref{sigma2}. Furthermore, there is no factor of 2 in front of $\epsilon^2$ since the uncertainty of the estimation of the opponents does not matter (because the team strength, averaged over all teams,  is zero). Finally, the factor $17^2$ expresses the fact that the team strength has to be expressed per 17 matches rather than per match. 

Without any information one obtains a standard deviation of 12 which, of course, is much larger than the optimum value of 6.8. The best observable is the packing rate, yielding a standard deviation of 7.5. However, there is still a lot of room for improvement which, however, is very difficult to achieve because appropriate observables need to be highly correlated with the team strength (in contrast to, e.g., the passes). As expected, just the knowledge of the results of the first half in terms of goals yields a relatively poor forecast (standard deviation of 8.05). 

\begin{table}[h!]
\caption{The parameters which are relevant to estimate the quality of the estimation of the goal difference of the second half of the season as well as the resulting standard deviation $(\lambda^2)^{0.5}$. The value of $A_X$ is taken from Ref.\cite{heuerarxiv}. See text for the definition of $A_X$ and $\epsilon^2$.}
\label{tab4}
      \begin{tabular}{llll}
        \hline
         X  & $A_X$ & $\epsilon^2$ & $\sqrt{\lambda^2}$   \\ \hline
       Goals & 0.82  & 0.118 & 8.85 \\
        Chances for goals & 0.90  & 0.068 & 8.05 \\
        Packing & 0.94 & 0.039 & 7.55\\
        Passes & 0.83 & 0.112 & 8.75\\
        no information & 0& 0.36& 12.0\\
      \hline
      \end{tabular}
\end{table}

\section{Discussion and Conclusion}

We have presented a systematic approach to disentangle informative and random
effects, describing soccer matches. For this purpose we have analysed the distribution of performance indicators as a function on the number of match days over which these performance indicators are averaged. The result, which is key for the statistical analysis, is expressed in Eq.\ref{varn2}.

Naturally, based on the availability of many match data (see, e.g., \cite{Pappalardo19}) a soccer match can be dissected in great detail and individual performance indicators can be analysed. However, what really matters is the final result. Therefore, we concentrated to a large extent on the goal difference, revealing new perspectives which may enable a deeper interpretation of soccer matches. Of course, the formalism can be applied to many different variables. We have also shown that the identification of informative and random effects and the respective interpretation of the different contributions to the random effects, also yield additional information.

As a key application, this information allows one to specify the optimum forecasting quality for a soccer match. It is given by the (on average) 20\% informative contribution to the overall variance of goal differences. Of course, the additional insight of the present work does not help to improve the forecasting of soccer matches but rather allows one to quantify its basic limits and therefore check, whether a given forecasting approach can still be significantly improved.

This discussion of randomness, naturally emerging from the statistical analysis in the present work, is very different to the emergence of special events in a match (e.g. hitting a bar or not).  Evidently, the whole soccer
match is a large collection of only imperfectly controlled actions.
Thus, one may say that top teams manage to control their
actions somewhat better. This does not reduce the
statistical randomness as expressed by $\delta^2$ but rather improves the team strength. Also a shot of Ronaldo from some distance may miss the goal significantly, but on average the likelihood to score a goal is higher than for most other players. This adds to the team strength but does not alter the notion of randomness, used in this work.

Naturally, beyond the random contributions the statistical analysis also allows one to estimate the informative contributions. We have shown one application of this value by discussing the stronger relevance of offensive actions as compared to defensive ones. We would like to stress that these results do not depend on a specific model of how soccer should work but rather are a consequence of the general statistical perspective.

We foresee that the present statistical framework can be used as well to obtain a more profound understanding about the information content of the myriad of data, generated during the extensive tracking of soccer matches and, of course, also beyond soccer.

\vspace{1cm}

{\bf Acknowledgement}

S. Reinartz and L. Keppler are acknowledged for supplying the data for the packing
rate and for helpful discussions.  Furthermore, I am grateful to Oliver
  Rubner for helpful discussions and technical support.



\end{document}